\documentclass[twoside]{article}

\usepackage{graphicx}

\usepackage{color}

\newcommand{\Tr}{\mathrm{Tr}}
\newcommand{\const}{\mathrm{const}}

\newcommand{\cM}{\mathcal{M}}
\newcommand{\cN}{\mathcal{N}}
\newcommand{\cT}{\mathcal{T}}

\newcommand{\dd}{\mathrm{d}}

\newcommand{\bC}{\mathbf{C}}

\begin{document}

% \title{}

% \author{Andrzej G\"orlich$^1$}

% \address{$^1$ Niels Bohr Institute, Blegdamsvej 17, 2100 Copenhagen, Denmark}

% \email{goerlich@nbi.dk}

\begin{center}
\vspace{24pt}
{ \large \bf The transfer matrix in four-dimensional Causal Dynamical Triangulations }\\[30pt]
{\sl A. G\"{o}rlich}\\[24pt]
{\footnotesize
The Niels Bohr Institute, Copenhagen University\\
Blegdamsvej 17, DK-2100 Copenhagen \O , Denmark.\\[5pt]
email: goerlich@nbi.dk
}
\end{center}

\begin{abstract}
\noindent Causal Dynamical Triangulations is a background independent approach to quantum gravity.
In this paper we introduce a phenomenological transfer matrix model,
where at each time step a reduced set of quantum states is used.
The states are solely characterized by the discretized spatial volume.
Using Monte Carlo simulations we determine the effective transfer matrix elements 
and extract the effective action for the scale factor. 
In this framework no degrees of freedom are frozen, however,
the obtained action agrees with the \emph{minisuperspace} model.
\end{abstract}

\section{Introduction}
The model of Causal Dynamical Triangulations (CDT) was proposed some years ago by J. Ambj\o rn, J. Jurkiewicz and R. Loll
with the aim of defining a lattice formulation of quantum gravity from first principles \cite{Dyna, Reco}.
The foundation of this model is the formalism of path-integrals applied to quantize a theory of gravitation.
The quantum gravity path integral is regularized 
by discretizing the spacetime geometry $g$ with piecewise linear manifold $\cT$.
The building blocks of four dimensional CDT are four-simplices,
which properly \emph{glued} along their faces form a simplicial manifold.

An important assumption of CDT is the causality condition.
As a consequence of the original Lorentzian signature of spacetime,
only causal geometries should contribute to the integral.
We will consider globally hyperbolic pseudo-Riemannian manifolds 
which allow introducing a global proper-time foliation.
The leaves of the foliation are spatial three-dimensional Cauchy surfaces called \emph{slices}.
Because topology changes of the spatial slices are often associated with causality violation,
we forbid the topology of the leaves to alter in time.
For simplicity, we chose 
the spatial slices to have a fixed topology of a three-sphere,
and establish periodic boundary conditions in the time direction.
Therefor, the spacetime topology is $\cM = S^1 \times S^3$.
Spatial slices of a triangulation are enumerated by a discrete \emph{time coordinate} $t$.
To each vertex of the triangulation such time coordinate is assigned,
bringing on a distinction between space-like links of length $a_s$ and time-like links of length $a_t$.
Because each simplex contains vertices lying in two consecutive spatial slices,
there are two kinds of simplices: first of a type $\{4, 1\}$ with four vertices lying in one spatial slice 
and one in the neighboring slice, and second of a type $\{3, 2\}$ with three vertices lying in one spatial slice and two in the adjacent slice.
The Wick rotation is performed by the analytic continuation to imaginary lengths of the time-like links $a_t \rightarrow i a_t$. 
The regularized partition function $Z$ is now written as a sum over causal triangulations $\cT$,
\begin{equation}
  Z = \int \mathcal{D}[g] e^{i S^{EH}[g]}
\ \rightarrow \ 
\sum_{\cT} e^{- S[\cT]} .
\label{eq:Partition}
\end{equation}
The Einstein-Hilbert action $S^{EH}[g] = \frac{1}{16 \pi G} \int \dd t \int \dd^3 x \sqrt{- g} (R - 2 \Lambda)$
evaluated on a simplicial manifold $\cT$ composed of $N_{4}$ simplices, among them $N_{41}$ being of type $\{4, 1\}$,
and $N_0$ vertices, gives the discrete Regge action, 
\begin{equation}
	S[\cT] = - K_0\ N_0 + K_4\ N_4  + \Delta\ (N_{41} - 6 N_0),
	\label{eq:SRegge}
\end{equation}
where $K_0$, $K_4$ and $\Delta$ are bare coupling constants,
and naively they are functions of $G, \lambda$ and $a_t, a_s$.

We applied Monte Carlo techniques,
and using the Regge action (\ref{eq:SRegge}), 
measured expectation values of observables within the CDT framework.
The simplest observable is the scale factor $a(t)$, 
or more conveniently the three-volume $n_t$
defined as the number of tetrahedra building slice $t$.

%%%% De Sitter phase %%%%
For a certain range of the coupling constants, 
a typical configuration is bell-shaped,
with the average volume profile 
$\langle n_t \rangle \propto \cos^3(t / B)$.
The emerged background geometry behaves like a well defined four-dimensional manifold
and is perfectly consistent with a Euclidean de Sitter universe, 
the classical vacuum solution of a spatially homogeneous and isotropic \emph{minisuperspace} model \cite{Background}.
In earlier work we have shown \cite{Plan} that the discretized \emph{minisuperspace} action,
\begin{equation}
	S[{n_t}] = \frac{1}{\Gamma} \sum_t \left( \frac{(n_{t+1} - n_t)^2}{n_{t+1} + n_t} + \mu n_t^{1/3} - \lambda  n_t  \right),
	\label{eq:SMiniD}
\end{equation}
describes well not only the measured $\langle n_t \rangle$ but also the fluctuations  
\begin{equation}
	\bC_{t t'} = \langle (n_t - \langle n_t \rangle) (n_{t'} -  \langle n_{t'}\rangle) \rangle .
	\label{eq:Ctt}
\end{equation}
The effective action (\ref{eq:SMiniD}) couples only adjacent slices.
Such form suggests that there exists an effective transfer matrix labeled only by the scale factor.

\section{The transfer matrix}

The model of Causal Dynamical Triangulations is completely determined by a transfer matrix $\cM$ 
labeled by three-dimensional triangulations $\tau$.
The matrix element $\langle \tau_1 | \cM | \tau_2 \rangle$  denotes the transition amplitude 
in one time step between states corresponding to triangulations $\tau_1$ and $\tau_2$. 
It is given by the sum over all four-dimensional triangulations $\cT$ of a slab, with boundary triangulations $\tau_1$ and $\tau_2$,
\[ \langle \tau_1 | \cM | \tau_2 \rangle = \sum_{\cT|_{\tau_1, \tau_2}} e^{-S[\cT] } .\]
The transfer matrix $\cM$ depends both on the entropy factor, 
which counts number of triangulations $\cT$ connecting the boundaries in one time step,
and the Regge action $S[\cT]$.
The partition function (\ref{eq:Partition}) corresponding to $T$ time steps is then expressed 
in terms of the matrix $\cM$,
\begin{equation}
	Z = \sum_{\mathcal{T}} e^{- S^{[\mathcal{T}]}} = \Tr \cM^T.
\label{eq:Z2}
\end{equation}
The probability of finding a configuration with $T$ spatial slices given by 
three-dimensional triangulations $\tau_1, \tau_2, \ldots, \tau_T$ is 
\begin{equation}
 P^{(T)} (\tau_1, \dots, \tau_T) = \frac{1}{Z} \langle \tau_1 | \cM | \tau_2 \rangle \langle \tau_2 | \cM | \tau_3 \rangle \dots  \langle \tau_T | \cM | \tau_1 \rangle .
\label{eq:Ptt}
\end{equation}

We used partition function (\ref{eq:Z2}) in Monte Carlo simulations.
The measurements performed so far, 
have been concentrated on the measurement of the three-volume $n_{t}$.
The probability $P^{(T)} (n_1, \dots, n_T)$ of finding a configuration with spatial volumes $n_1, n_2, \ldots, n_T$ 
is given by a proper sum of partial probabilities (\ref{eq:Ptt}).
Let $T_3(n)$ denote the subset of three-dimensional triangulations which are build of exactly $n$ three-simplices.
We use the projection operator $\rho(n) \equiv | n \rangle\langle n |$ on the subspace spanned by $T_3(n)$,
\begin{equation}
\rho(n) \equiv | n \rangle\langle n | \equiv \sum_{\tau \in T_3(n)} | \tau \rangle\langle \tau |.
\label{eq:rhon}
\end{equation}
to express the probability $P^{(T)} (n_1, \dots, n_T)$,
% Expression (\ref{eq:PTnP}) is simplified by inserting $| n \rangle\langle n |$,
\begin{equation}
	P^{(T)} (n_1, \dots, n_T) = \frac{1}{Z}  \Tr \left[| n_1 \rangle\langle n_1 | \cM | n_2 \rangle \langle n_2 | \cM | n_3 \rangle \dots  \langle n_T | \cM \right].
\label{eq:PTncM}
\end{equation}
In (\ref{eq:PTncM}) it is misleading to think of the aggregated ``state'' $| n \rangle$ as
a normalized sum of the vectors $|\tau\rangle,\ \tau \in T_3(n)$.
Such a vector would again be a single vector located in the 
space spanned by the $|\tau \rangle$'s.
It is more appropriate to interpret the ``state'' associated with $n$ 
as arising from a classical uniform {\it probability distribution} of states
$|\tau \rangle$  and in this way to treat $\rho(n)$ as the associated
{\it density operator}. 

As mentioned in the Introduction,
the form of the effective action (\ref{eq:SMiniD})
obtained from the covariance matrix (\ref{eq:Ctt})
suggests that there exists an effective transfer matrix $\langle n | M | m \rangle$
whose elements are labeled by the three-volumes
and that it is possible to effectively decompose
observed distributions $P^{(T)} (n_1, \dots, n_T)$ into a product
\begin{equation}
	P^{(T)} (n_1, \dots, n_T) = \frac{1}{Z} \langle n_1 | M | n_2 \rangle \langle n_2 | M | n_3 \rangle \cdots \langle n_T | M | n_1 \rangle .
\label{eq:PTnM}
\end{equation}
The effective transfer matrix $M$ depends only on the coupling constants $K_0$, $\Delta$ and $K_4$
but not on the number of slices $T$.
In analogy to (\ref{eq:Ptt}), the elements of the effective transfer matrix correspond to 
transition amplitudes in one time step between states of a given three-volume.

\section{Measurements}

In the following, we will assume that we can work with an 
effective transfer matrix $\langle n|M|m \rangle$
and will show that equation (\ref{eq:PTnM}) 
provides a very good approximation of measured data \cite{Transfer}.

For simplicity, let us define the two-point function,
\begin{equation}
	P^{(T)} (n_t, n_{t + \Delta t}) = \frac{1}{Z} \langle n_t | M^{\Delta t} | n_{t + \Delta t} \rangle \langle n_{t + \Delta t} | M^{T - \Delta t} n_t \rangle .
\label{eq:PTwoPoint}
\end{equation}
by summing (\ref{eq:PTnM}) over all three-volumes except for times $t$ and $t + \Delta t$.
The simplest way to measure the matrix elements $\langle n | M | m \rangle$, up to a normalization,
is to consider $T = 2$,
\[ \langle n | M | m \rangle \propto \sqrt{P^{(2)} (n_1 = n, n_2 = m)}.\]

The effective transfer matrix elements can be measured in various ways.
In particular, as follows from (\ref{eq:PTwoPoint}) for $T = 3, 4$,
we have
\begin{equation}
\langle n | M | m \rangle \propto \frac{P^{(3)}(n_1 = n, n_2 = m)}{\sqrt{P^{(4)}(n_1 = n, n_3 = m)}} .
\label{eq:TM34}
\end{equation}
We tested, that the elements $\langle n | M | m \rangle$ 
measured in different ways completely agreed up to numerical noise,
supporting validity of equation (\ref{eq:PTnM}).
For technical reasons, most measurements were performed using expression (\ref{eq:TM34}).

\begin{figure}
\begin{center}
\includegraphics[width=0.48\textwidth]{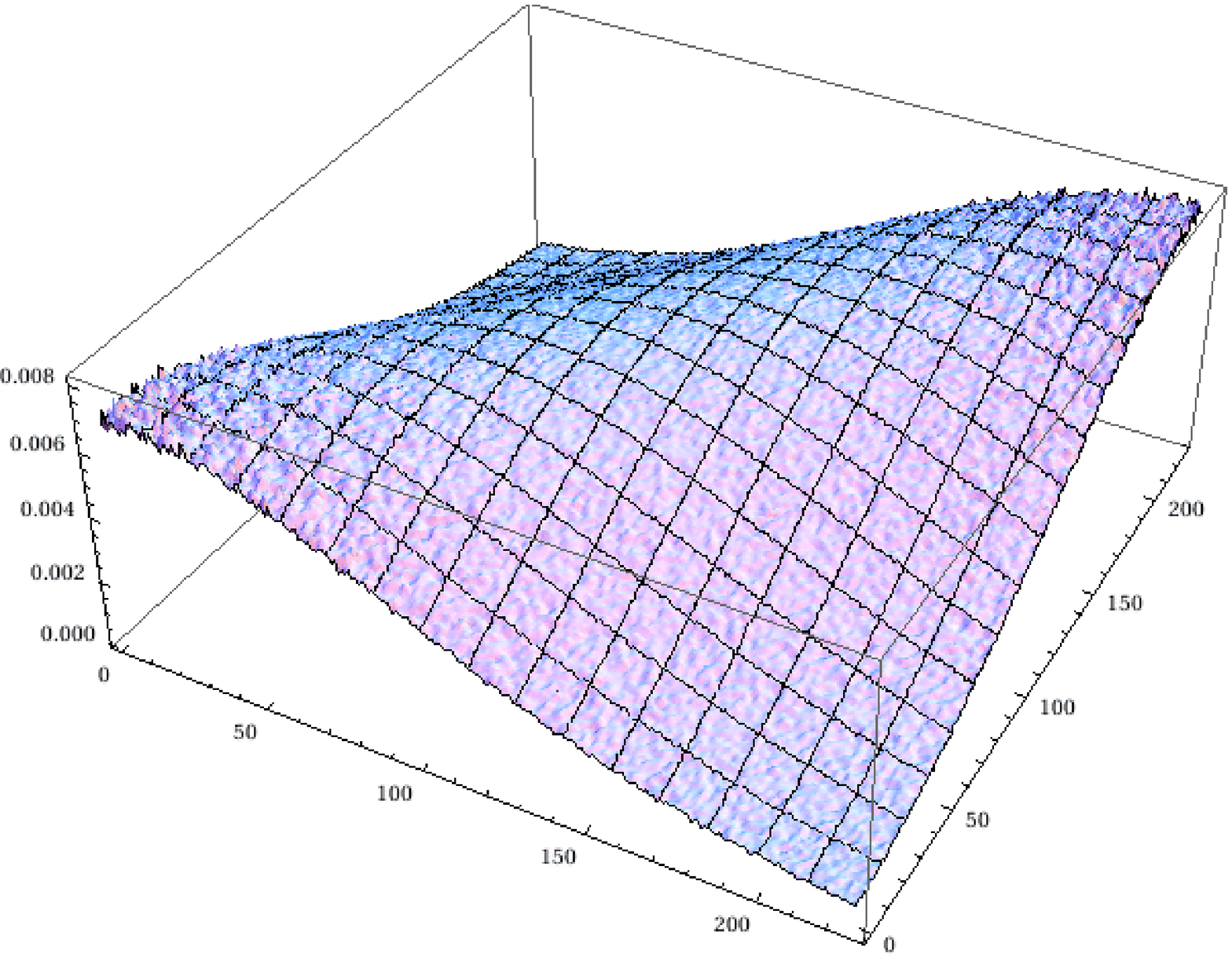}
\includegraphics[width=0.48\textwidth]{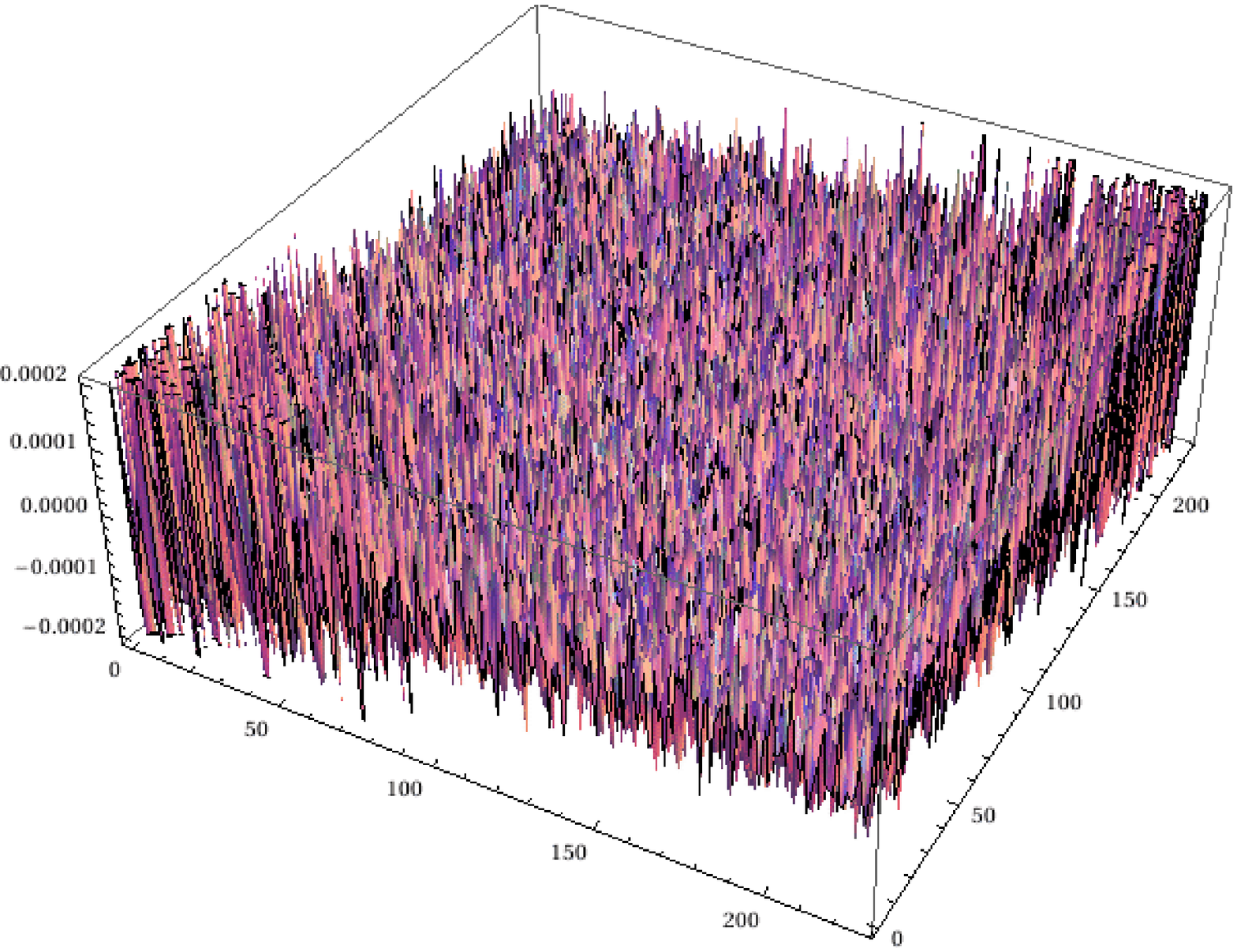}
\end{center}
\caption{
Left: The \emph{empirical} transfer matrix for range $1200  < n_t < 1600$.
Right: The difference between the empirical and theoretical matrices 
disappears in the numerical noise.
}
\label{fig:TMXL}
\end{figure}

The coupling constant $K_4$ in (\ref{eq:SRegge}) plays a role of a cosmological constant.
To correctly perform simulations, we have to approach with $K_4$ very close to its critical value $K_4^{crit}$. 
To efficiently probe desired range of the three-volume, we added to the Regge action (\ref{eq:SRegge})
a quadratic term to fix $n_t$ around $n_{vol}$,
\[ S \rightarrow S + \epsilon \sum_t (n_{t} - n_{vol})^2 . \]
Because it is consistent with the decomposition (\ref{eq:PTnM}), 
its effect can be easily canceled.
For technical reasons, we measured the transfer matrix $M$ 
separately for few overlapping ranges of the three-volume.

\section{The effective action}

The \emph{effective} action obtained from the covariance matrix (\ref{eq:Ctt})
is directly related to the \emph{effective} transfer matrix $M$.
The \emph{minisuperspace} action (\ref{eq:SMiniD}) suggests that the \emph{effective} transfer matrix given by
\begin{equation}
\langle n | M | m \rangle = \cN e^{-\frac{1}{\Gamma} \left[ \frac{(n - m)^2}{n + m} + \mu \left( \frac{n + m}{2} \right)^{1/3} -\lambda \frac{n + m}{2} \right]} 
\label{eq:tmeff}
\end{equation}
is a good approximation in the bulk where $n_t$ is large
\footnote{We slightly modified the form of the potential term. Such parametrization is more convenient to extract the parameters of the action.}.
Further, we will measure the empirical transfer matrix elements $\langle n | M | m \rangle$,
extract the parameters $\Gamma$, $\mu$ and $\lambda$, and check that (\ref{eq:tmeff})
is indeed a good approximation of the data.
The measured \emph{effective} transfer matrix $M$,
for range $1200  < n_t < 1600$,
is presented in Fig. \ref{fig:TMXL} (left graph).
The right graph shows the difference between the measured matrix $M$ and the best fit (\ref{eq:tmeff}).
Indeed, the difference disappears in the numerical noise proving that the approximation (\ref{eq:tmeff}) is very good.

The measurements presented in this paper were performed for coupling constants $K_0 = 2.2$, $\Delta = 0.6$ and $K_4 = 0.922$.

\subsection{The kinetic term}

\begin{figure}
\begin{center}
\includegraphics[width=0.48\textwidth]{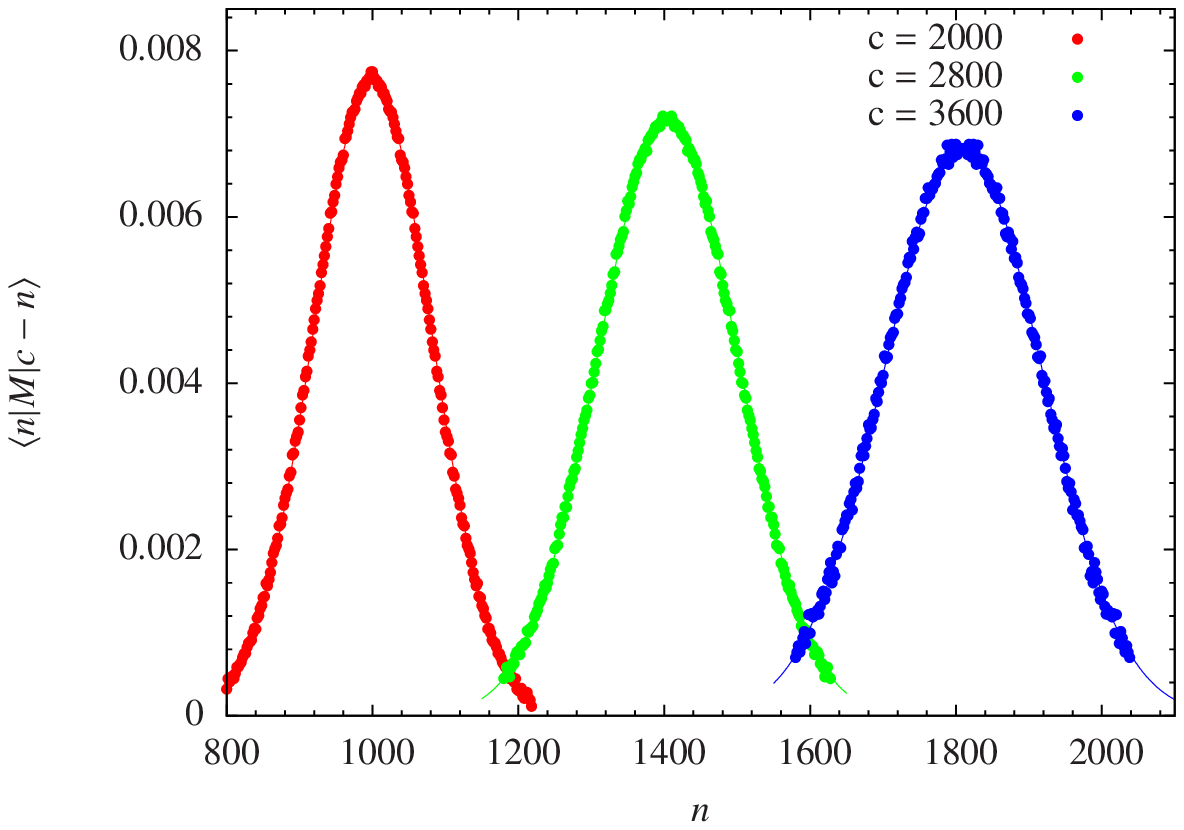}
\includegraphics[width=0.48\textwidth]{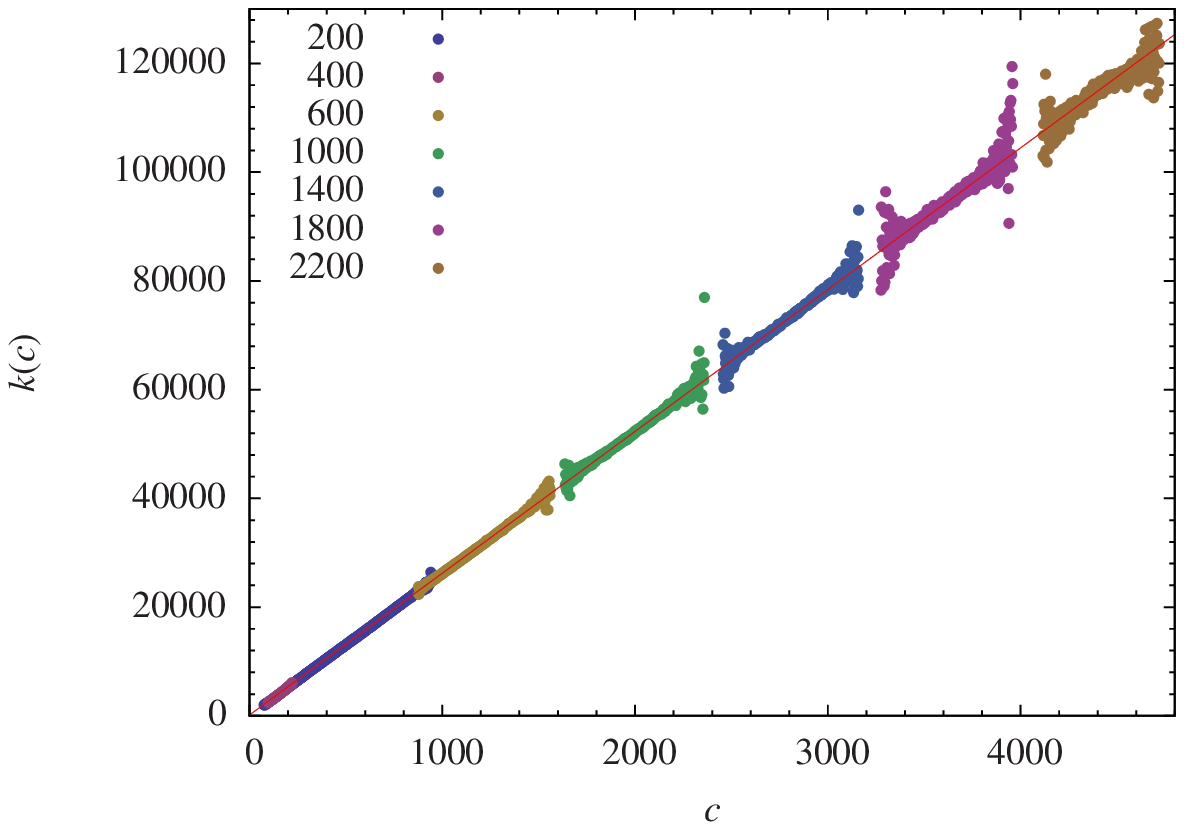}
\end{center}
\caption{
Left: $\langle n | M | c - n \rangle$ plotted as a function of $n$ for various $c$ (dots).
Gaussian fits are drawn with a line.
Right: the coefficient $k(c)$ in the kinetic term as a function of $c = n + m$,
(different colors denote different ranges) and a linear fit $k(n + m) = \Gamma \cdot (n + m)$ (red line).}
\label{fig:kin}
\end{figure}

To get a better estimation of the parameters associated with
the effective action (\ref{eq:SMiniD}) and (\ref{eq:tmeff}),
we first try to fit only to the parameters of the kinetic term 
which is the dominating term from a numerical point of view. 
We do that by keeping the sum of the entries, i.e. $n+m$, fixed
such that the potential term is not changing. 
In this way we determine $\Gamma$ with high accuracy.
The matrix elements for constant $n + m = c$
show the expected Gaussian dependence on $n$ 
(see left Fig. \ref{fig:kin}),
\begin{equation}
\langle n | M | m \rangle = \langle n | M | c - n \rangle = 
\mathcal{N}(c) \exp \left[- \frac{(2 n - c)^2}{\Gamma \cdot c} \right], 
\end{equation}
where the terms in the effective action which only depend on $c$ are 
included in the normalization.

We expect the denominator of the kinetic term $k(c)$ to behave like 
$k(n + m) = \Gamma \cdot (n + m)$.
As shown in the right graph in Fig. \ref{fig:kin}
this is indeed true and the parameter $\Gamma$ 
is constant in the whole range of the three-volumes.
% Fig. \ref{FigKinetic} presents measured 
% coefficients $k(c)$ for various $c$'s and ranges of $n_t$
% denoted by distinct colors together with a linear fit.
The best linear fit gives $\Gamma = 26.07 \pm 0.05$.
% This result is consistent with the fits obtained for separate ranges of $n_t$.
This result is consistent with the values obtained from the fits for separate ranges of $n_t$.

\subsection{The potential term}

\begin{figure}
\begin{center}
\includegraphics[width=0.8\textwidth]{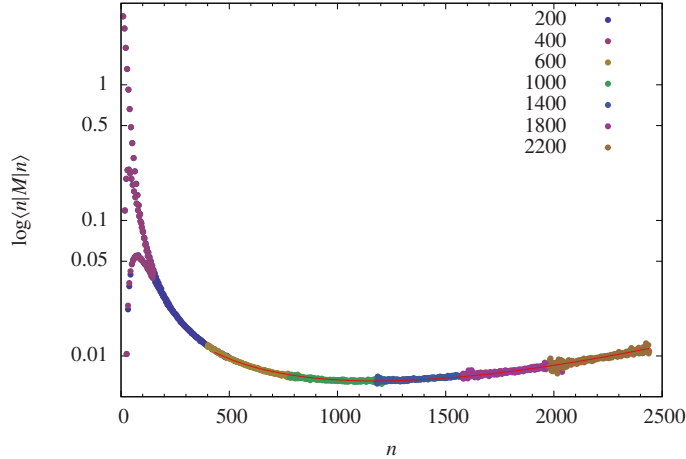}
\end{center}
\caption{
$\log \langle n | M | n \rangle$ of the scaled transfer matrix 
(dots, different colors denote different ranges)
compared with the fit of the potential term $-L_{eff}$ (red line,
which stops at $n=400$).
}
\label{fig:pot}
\end{figure}

The potential part of the effective Lagrangian $L_{eff}$
may be extracted from the diagonal elements of the transfer matrix,
\begin{equation}
 L_{eff}(n, n) = - \log \langle n | M | n \rangle + \const = 
\frac{1}{\Gamma} \left( \mu n^{1/3} - \lambda n \right).
 \label{eq:Leff}
\end{equation}

For technical reasons, we measured the transfer matrix $M$ 
separately for few different ranges of the three-volume.
Because, the normalization is not uniquely defined,
in order to merge the effective Lagrangian,
the constant in (\ref{eq:Leff}) has to be properly adjusted.
The measured merged effective Lagrangian is shown in Fig.\ \ref{fig:pot}.
The colors denote different ranges for which the transfer matrix was measured.
Fig.\ \ref{fig:pot} presents also the fit of form (\ref{eq:Leff}).
In the bulk region, where $n_t$ is large enough, the theoretical expectation (\ref{eq:Leff}) fits very well.
The measured values are $\mu = 16.5 \pm 0.2$ and $\lambda = 0.049 \pm 0.001$, where we took $\Gamma = 26.07$.
Again, this result is consistent with the values obtained from the fits for separate ranges of $n_t$.

\section{Conclusions}

The model of Causal Dynamical Triangulations comes with a transfer matrix $\langle \tau_1 | \cM | \tau_2 \rangle$.
The measured distributions of the three-volumes $n_t$, e.g. $P^{(T)}(n_t, n_{t + \Delta t})$, 
have an exact definition in terms of the \emph{full} transfer matrix $\cM$ and the density matrix $| n \rangle \langle  n|$.
The actual data coming from Monte Carlo simulations seem to allow for a much simpler description 
in terms of an \emph{effective} transfer matrix $M$,
labeled by \emph{abstract} vectors $| n \rangle$ referring only to the three-volume.
The \emph{effective} transfer matrix $M$ allows to directly measure the \emph{effective} action $S[n_t]$.
An important advantage of the present method, since number of slices $T$ is small,
is much faster measurement of the transfer matrix 
compared to the covariance matrix, which was used previously to extract the effective action.
Basically over the whole range of $n_t$ the \emph{effective} transfer matrix elements can be represented as
\[ \langle n | M | m \rangle = \cN e^{-\frac{1}{\Gamma} \left[ \frac{(n - m)^2}{n + m} + \mu \left( \frac{n + m}{2} \right)^{1/3} -\lambda \frac{n + m}{2} \right]} , \]
with high accuracy. 
This result is fully consistent with the reduced \emph{minisuperspace} action (\ref{eq:SMiniD}),
although in CDT we do not freeze any degrees of freedom.

An issue not addressed in this article,
is the problem of small three-volumes.
For small $n_t$ we do not observe a Gaussian distribution 
of the three-volume $n_t$ around the mean value $\langle n_t \rangle$.
Because of strong discretization effects,
the probability distributions, and consequently the \emph{effective} transfer matrix elements,
split into three families \cite{Semiclassical}.
Despite different nature, after the \emph{smoothing} procedure,
the effective action for small volumes is basically the same as for large volumes,
with a small modification in the potential \cite{Transfer}.
It might be interpreted as possible curvature corrections,
however, we are not able to measure it accurately in a discretization independent way.

\vspace{1cm}
\noindent
{\bf Acknowledgments}\\
The author acknowledges support by the Danish Research Council 
grant ``Quantum gravity and the role of Black holes''.


\begin{thebibliography}{99}


\bibitem{Dyna}
 J. Ambj{\o}rn, J. Jurkiewicz and R. Loll,
 {\it Dynamically triangulating Lorentzian quantum gravity},
 Nucl. Phys. {\bf B610}, 347, (2001) [hep-th/0105267].

\bibitem{Reco}
 J. Ambj{\o}rn, J. Jurkiewicz and R. Loll,
 {\it Reconstructing the universe},
 Phys. Rev. {\bf D72}, 064014, (2005) [hep-th/0505154].

\bibitem{Plan}
 J. Ambj{\o}rn, A. G\"orlich, J. Jurkiewicz and R. Loll,
 {\it Planckian Birth of the Quantum de Sitter Universe},
 Phys. Rev. Lett. {\bf 100}, 091304, (2008) [arXiv:0712.2485].

\bibitem{Background}
 A. G\"orlich,
 {\it Background Geometry in 4D Causal Dynamical Triangulations},
 Acta Phys. Pol. {\bf B39}, 3343, (2008).

\bibitem{Transfer}
 J. Ambj{\o}rn, J. Gizbert-Studnicki, A. G\"orlich and J. Jurkiewicz,
 {\it The transfer matrix in four-dimensional CDT},
 JHEP {\bf 1209}, 017, 2012 [arXiv:1205.3791].

\bibitem{Semiclassical}
 J. Ambj{\o}rn, A. G\"orlich, J. Jurkiewicz, R. Loll, J. Gizbert-Studnicki and T. Trze\'sniewski,
 {\it The semiclassical limit of Causal Dynamical Triangulations},
 Nucl. Phys. {\bf B849}, 144, (2011) [arXiv:1102.3929].

\end{thebibliography}
\end{document}